\documentclass[twocolumn, trackchanges]{aastex61}

\usepackage[utf8]{inputenc} 
\usepackage{natbib}

\usepackage{graphicx} 
\usepackage{amsbsy}
\usepackage{bm}
\received{January 25, 2018}
\revised{February 15, 2018}
\accepted{April 25, 2018}
\published{--}
\submitjournal{The Astrophysical Journal}

\shorttitle{Kinetic MRI turbulence}
\shortauthors{Inchingolo et al.}

\begin{document}

\title{Fully kinetic large scale simulations of the collisionless Magnetorotational instability}

\author{Giannandrea Inchingolo}
\affiliation{GoLP/Instituto de Plasmas e Fusão Nuclear, Instituto Superior T\'ecnico, Universidade de Lisboa, 1049-001 Lisboa, Portugal}
\affiliation{Plasma Science and Fusion Center, Massachusetts Institute of Technology, Cambridge, MA 02139, USA}

\author{Thomas Grismayer}
\affiliation{GoLP/Instituto de Plasmas e Fusão Nuclear, Instituto Superior T\'ecnico, Universidade de Lisboa, 1049-001 Lisboa, Portugal}

\author{Nuno F. Loureiro}
\affiliation{Plasma Science and Fusion Center, Massachusetts Institute of Technology, Cambridge, MA 02139, USA}

\author{Ricardo A. Fonseca}
\affiliation{GoLP/Instituto de Plasmas e Fusão Nuclear, Instituto Superior T\'ecnico, Universidade de Lisboa, 1049-001 Lisboa, Portugal}
\affiliation{DCTI/ISCTE Instituto Universitário de Lisboa, 1649-026 Lisboa, Portugal}

\author{Luis O. Silva}
\affiliation{GoLP/Instituto de Plasmas e Fusão Nuclear, Instituto Superior T\'ecnico, Universidade de Lisboa, 1049-001 Lisboa, Portugal}

\begin{abstract}
We present two-dimensional particle-in-cell (PIC) simulations of the fully kinetic collisionless magnetorotational instability (MRI) in weakly magnetized (high $\beta$) pair plasma.
The central result of this numerical analysis is the emergence of a self-induced turbulent regime in the saturation state of the collisionless MRI, which can only be captured for large enough simulation domains.
One of the underlying mechanisms for the development of this turbulent state is the drift-kink instability (DKI) of the current sheets resulting from the nonlinear evolution of the channel modes.
The onset of the DKI can only be observed for simulation domain sizes exceeding several linear MRI wavelengths.
The DKI, together with ensuing magnetic reconnection, activate the turbulent motion of the plasma in the late stage of the nonlinear evolution of the MRI.
At steady state, the magnetic energy has an MHD-like spectrum with a slope of $k^{-5/3}$ for $k\rho<1$ and $k^{-3}$ for sub-Larmor scale ($k\rho>1$).
We also examine the role of the collisionless MRI and associated magnetic reconnection in the  development of pressure anisotropy.
We study the stability of the system due to this pressure anisotropy, observing the development of mirror instability during the early-stage of the MRI.
We further discuss the importance of magnetic reconnection for particle acceleration during the turbulence regime.
In particular, consistent with reconnection studies, we show that at late times the kinetic energy presents a characteristic slope of $\epsilon^{-2}$ in the high-energy region.
\end{abstract}
 

\section{Introduction} \label{sec:intro}
Accretion disks are astrophysical structures in which a gas or a plasma rotates around a massive central object, such as a black hole or a neutron star, under the effect of the gravitational force~\citep{Pringle1981}. In particular, since the gravitational force decreases as the distance from the central object increases, the angular velocity of the plasma is, therefore, lower far from the central object. 
This property of accretion disks induces the development of the so-called magnetorotational instability (MRI) \citep{Chandrasekhar1958, Balbus1991}, through the action of which an initial seed (weak) magnetic field is exponentially amplified on a time scale comparable with the typical rotational period of the disk. 
Theoretical arguments and numerical simulations suggest that the saturation amplitude of the MRI is such that there is an approximate equipartition between the kinetic and magnetic energies (i.e., the plasma $\beta$, initially very large, saturates at values around 1)~\citep{Hawley1995,Balbus1998}.

The current understanding of the MRI stems largely from MHD theory and simulation~\citep{Balbus1998}. \cite{Goodman1994} discussed the formation of large-scale coherent structures --- channel flows --- in the early nonlinear regime of the MRI.
These structures have been shown to be unstable to parasitic instabilities:  
Kelvin-Helmholtz, tearing, kink and pinch have all been suggested as possible modes that may play a crucial role in the disruption of the channel flows and subsequent activation of a turbulent stage~\citep{Goodman1994,Pessah2009,Latter2009}.

In radiatively inefficient accretion flow models for accretion onto compact objects, the accretion proceeds via a hot, low-density plasma with the proton temperature larger than the electron temperature (see \cite{Narayan1998} and \cite{Quataert2003} for reviews).
In order to maintain such a two-temperature flow, the typical collision rate must be much smaller than the accretion rate.
This suggests that the standard MHD approach for the description of the dynamics of such accretion disks may be insufficient, and a kinetic description is required instead. 
Indeed, several theoretical studies of collisionless MRI (e.g. \cite{Quataert2002, Sharma2003, Krolik2006, Sharma2006, Sharma2007}) have shown the development of pressure anisotropies during the evolution of the MRI when kinetic effects are taken into account. 
Fundamentally, this is due to the fact that in typical accretion disks, the growth rate of the MRI is much smaller than the ion cyclotron frequency, and so the magnetic moment $\mu = mv_\perp^2/2B$, where $v_\perp$ is the component of the velocity perpendicular to the magnetic field $B$, ought to be conserved. 
The amplification of the magnetic field produced by the MRI therefore leads to an increase in $v_\perp$. 
The absence of significant collisions implies that $v_\perp$ and $v_\parallel$ will thus become different, originating a pressure anisotropy.
Several studies have shown that this anisotropy activates various kinetic instabilities, including, e.g., the mirror and the firehose~\citep{Rosenbluth1956, Parker1958, Hasegawa1969, Yoon1993, Gary1993, Gary1997, Pokhotelov2000, Hellinger2000, Pokhotelov2004, Gary2006, Kunz2014, Kunz2015a}.
It is conjectured that these instabilities may significantly affect the nonlinear development of the MRI, and critically impact the transport of momentum and energy in accretion disks~\citep{Mogavero2014,Kunz2014,Melville2016,Kunz2016}.

The generation of a pressure anisotropy is not exclusive to the ions; indeed, it is expected that electrons will also develop a non-unity ratio of $v_\perp$ and $v_\parallel$, and thus trigger their own pressure anisotropy instabilities~\citep{Gary2006}. Their effect on MRI development and saturation, and the interplay between ion and electron scale instabilities, is not currently understood. 

Addressing these questions requires first-principles, fully kinetic simulations --- as does the detailed understanding of energy partition and dissipation.
Unfortunately,  the typical range of scales and frequencies of the collisionless MRI is such that global kinetic simulations of accretion disks are impossible with present day computational resources. 
However, the simulation of a local portion of the disk (the local shearing-box approximation) is just about feasible, and might be sufficient to gain insight into how the collisionless MRI behaves. 
In their pioneering PIC numerical studies, using a shearing co-rotating framework, Riquelme {\it et al.} showed that the MRI generates pressure anisotropies in both low \citep{Riquelme2012} and high $\beta$ \citep{Riquelme2015} regimes, and argued that to be the reason for their observation of the mirror instability. 
Nonetheless, those simulations did not reach the saturation stage of the MRI within the limits of validity of the shearing co-rotating model used. 
Using a co-rotating framework with shearing periodic boundary conditions, Hoshino performed two-dimensional \citep{Hoshino2013} and three-dimensional \citep{Hoshino2015} PIC simulations of electron-positron plasma for weakly magnetized, non-relativistic, collisionless MRI with  $\beta\gg 100$, confirming the generation of the mirror instability in the linear regime of the MRI and emphasizing the role of magnetic reconnection in the saturation of the MRI for the high $\beta$ regime.
Due to the large computational cost of these studies, Hoshino's simulations were limited to the analysis of relatively small simulation domains.
Despite yielding saturation of the MRI, these simulations did not reach a turbulent stage.

Observing the saturation stage of the MRI with realistic ion-electron mass ratios and sufficiently large simulation domains is presently an insurmountable challenge for PIC simulations. 
A compromise can be found by using hybrid-kinetic codes, which treat the ions kinetically, but retain a fluid description of the electrons. 
Using this approach, \cite{Kunz2016} demonstrated that the saturation of the kinetic MRI proceeds via a steady-state turbulent regime.

This work presents an investigation that is complementary to these studies.
We perform \textit{ab-initio} two-dimensional PIC simulations of collisionless MRI in a pair plasma. 
The PIC description of both species intrinsically includes all kinetic effects; but our study misses potential effects that critically depend on dimensionality and scale separation between the two species.
Such a compromise is imposed by the rather stringent computational limitations that characterize  this problem.
We investigate the dependence of our results on the size of simulation domain and observe, for the first time in fully kinetic studies, the turbulent saturation of the MRI, provided that the simulation domain is sufficiently large compared to the wavelength of the linearly most unstable MRI mode. 
When that is the case, we witness the onset of the drift-kink instability (DKI) in the nonlinear regime. The combined effect of this instability with magnetic reconnection of the channel flows, appears to be the key ingredients to trigger a turbulent regime.
 
This paper is organized as follows.
In Section \ref{sec:model} we describe the shearing co-rotating framework implemented in our numerical code \software{OSIRIS \citep{Fonseca2002, Fonseca2013}}.
The nonlinear evolution of the MRI is analyzed in Section \ref{sec:box_size} describing the generation of the drift-kink instability in our larger simulations and how this influences the generation of subsequent turbulent regime in the plasma.
Section \ref{sec:mirror} focuses on the generation of pressure anisotropy and the consequent mirror instability.
In section \ref{sec:turbo} we investigate the turbulent saturation regime; the detailed analysis of the energy distribution and particle acceleration in this stage is reported in Section \ref{sec:acceleration}. 
Section \ref{sec:conclusion} summarizes the results obtained in the paper and discusses  future work.

\section{Shearing co-rotating frame} \label{sec:model}
Our numerical calculations employ the shearing co-rotating frame developed by \cite{Riquelme2012}. For completeness, we summarize here the key aspects of this model; the reader is referred to that reference for details.

In order to study the evolution of the collisionless accretion disk, we investigate the two-dimensional (poloidal) $x-z$ plane, where $x$ is the radial direction of the accretion disk and $z$ the vertical direction, parallel to the rotation axis. 
The $y$ direction, perpendicular to the simulation plane, represents the transverse (toroidal) direction of the accretion disk. 
At equilibrium, the accretion disk follows Keplerian orbits around the central mass, where the plasma at each radial position $x_0$ rotates with angular velocity $\pmb{\Omega}=\Omega(x_0)\hat{z}\propto1/x_0^2$. 
In our simulation, we study the evolution of a local portion of the disk, centered around the equilibrium position $x_0$, and require that the simulation domain in the radial direction, $L_x$, be small compared to the equilibrium position $L_x\ll x_0$. 
In this approximation, the shearing velocity of the Keplerian disk, $\bf{v_0} = \pmb{\Omega}\times\bf{x_0}$, can be linearized to $\bf{v_0}$ $= -3/2\alpha x\hat{y}$, where $\alpha=(d\Omega/dx)|_{x_0}$.

In order to include a differential shearing velocity $\bf{v_0}$, the standard approach, used both in MHD simulations (see for example \cite{Hawley1995}) and kinetic simulations \citep{Hoshino2013, Hoshino2015, Kunz2014}, consists in the implementation of shearing periodic boundary conditions along the radial direction. 
An alternative method has been proposed by \cite{Riquelme2012}, whereby one performs a Galilean transformation of equations (\ref{eq:gauss}-\ref{eq:pusher}) implementing shearing coordinates.

To conduct our numerical simulations, we follow this last approach, modifying the PIC code \software{OSIRIS \citep{Fonseca2002, Fonseca2013}} to include a local shearing co-rotating framework.
In this particular frame, the shearing term $\bf{v_0}$ appears explicitly in the equations.
The implementation of the shearing co-rotating frame requires a series of approximations that we now discuss.
The co-rotating reference frame is non-inertial, so Maxwell's equations become \citep{Schiff1939}
\begin{eqnarray}
   \nabla\cdot\bm{E} & = & 4\pi\rho+\frac{2\pmb{\alpha}\cdot\bm{B}}{c}-\frac{\bm{v_0}}{c}\cdot\nabla\times\bm{B}, \label{eq:gauss}\\
    \nabla\cdot\bm{B} & = & 0, \label{eq:gradb}\\
    \frac{\partial\bm{B}}{\partial t} & = & -c\nabla\times\bm{E}, \label{eq:faraday}\\
    \frac{\partial\bm{E}}{\partial t} & = & c\nabla\times\bm{B} - 4 \pi\bm{J} + \frac{\bm{v_0}}{c}\times\frac{\partial\bm{B}}{\partial t} \nonumber\\
    & & -\nabla\times\bigg(\bm{v_0}\times\bigg(\bm{E}-\frac{\bm{v_0}}{c}\times\bm{B}\bigg)\bigg), \label{eq:ampere}
\end{eqnarray}
where $\pmb{\alpha} =\alpha\hat{z}$ is the angular frequency of the accretion disk.
In order to simplify these equations, we limit our analysis to the non-relativistic case in which we can neglect the last two terms in Equation (\ref{eq:gauss}) and all terms proportional to $v_0$ in Equation (\ref{eq:ampere}).
Strictly speaking, the non-relativistic approximation would also require us to neglect the displacement current. 
However, the details of the PIC numerical algorithm require us to keep it to update the electric field. There is no inconsistency between keeping the displacement current and neglecting the terms proportional to $v_0$ provided that the equilibrium position $x_0$ is chosen to be large enough.
With the non-relativistic approximation, the Maxwell's equations (\ref{eq:gauss}-\ref{eq:ampere}) simplify to the usual ones.

In the co-rotating frame, the motion of the plasma is affected by the Coriolis force. In the case of a Keplerian disk, this is given by the well-known expression
\begin{equation}
    \frac{d \bm{p}}{dt}=q\bigg(\bm{E}+\frac{\bm{v}\times\bm{B}}{c}\bigg)-2\pmb{\alpha}\times\bm{p},\label{eq:pusher}
\end{equation}
where $\bm{p}$ and $\bm{v}$ are the particle momentum and velocity and $q$ is its charge. 
The expression for the Coriolis force is valid in the cold limit, where the fluid velocity $|\bm{u}|$ is small compared to the shear velocity, $|\bm{u}|\ll |\bm{v_0}|$. 

The equation of motion (\ref{eq:pusher}) is valid in any non-relativistic co-rotating frame.
The non-relativistic restriction imposed above for the simplification of the Maxwell's (\ref{eq:gauss}-\ref{eq:ampere}) and momentum (\ref{eq:pusher}) equations refers to the main bulk velocity of the plasma.
There is no restriction on the motion of single particles, which can in principle be relativistic.
The set of equations for the non-relativistic case remains valid also in the presence of relativistic particles, as long as the fluid motion of the plasma remains non-relativistic.
For further discussion on the neglected relativistic effects in a generic co-rotating frame, we refer the reader to \cite{Riquelme2012}.

To move from the co-rotating frame just described -- in the non-relativistic limit -- to the final shearing co-rotating frame, we apply a Galilean transformation, following the approach described in the Appendix of \cite{Riquelme2012}.
The shearing, co-rotating Maxwell's equations become
\begin{eqnarray}
	\nabla\cdot\bm{B} & = & 0, 
	\label{eq:divB}\\
    \nabla\cdot\bm{E} & = & 4\pi\rho, \label{eq:three-dimensionshear_gauss}\\
    \frac{\partial\bm{B}}{\partial t} & = & -c\nabla\times\bm{E} - \frac{3}{2} \alpha B_x\hat{y}, \label{eq:three-dimensionshear_faraday} \\
    \frac{\partial\bm{E}}{\partial t} & = & c\nabla\times\bm{B}-4\pi\bm{J} -\frac{3}{2}\alpha E_x\hat{y}, \label{eq:three-dimensionshear_ampere}
\end{eqnarray}
and the equation of motion transforms to
\begin{equation}
    \frac{d \bm{p}}{dt}=q\bigg(\bm{E}+\frac{\bm{v}\times\bm{B}}{c}\bigg)-2\pmb{\alpha}\times\bm{p}+\frac{3}{2}\alpha p_x\hat{y},\label{eq:shear_pusher}
\end{equation}
where we neglected the terms proportional to $\partial/\partial y$ in Eq. (\ref{eq:divB}-\ref{eq:three-dimensionshear_ampere}) since we restrict our study to two-dimensions (the poloidal ($x$-$z$) plane).
As we present our results below, we will attempt to discuss how the extension to a three-dimensional setup might affect them, or not.

To verify the validity of the results obtained in the two-dimensional shearing co-rotating frame, we performed benchmarks against the linear theory of collisionless MRI~\citep{Krolik2006}, adapted for pair plasmas; this is reported in Appendix \ref{sec:appendix}.

\subsection{Simulation setup}
We start with a non-relativistic, isotropic, weakly magnetized pair plasma ($e_+-e_-$) with $\beta=8\pi(p_+ + p_-)/B^2_0=100$, where the pressure of each species is related to their respective thermal velocity $v_{th,\pm}=(3k_BT_\pm/m)^{1/2}$ by $ p_\pm = (1/2)m nv^2_{th,\pm}$.
As discussed in the Introduction, the choice of a pair plasma enables us to simulate several orbital periods ($2\pi/\alpha$) of the accretion disk and much larger simulation domains  that would not be possible with realistic mass ratios.

The external magnetic field is set to be vertical to the accretion disk, i.e., $\bm{B_0}=$ $B_0\hat{z}$.
Its initial value is set using the corresponding Alfv\'en speed $v_{A,0}=B_0/\sqrt{4\pi m n}$ and fixed to $v_{A,0}/c=1.43\times10^{-2}$. 
The orbital frequency $\alpha$ is expressed in terms of the initial cyclotron frequency $\Omega_0=eB_0/mc$ and fixed to $\alpha/\Omega_0 = 1/11$. 
The simulation domains $L_x$ and $L_z$ are normalized by $\lambda_0=2\pi v_A/\alpha$ (approximately the wavelength of the fastest growing MRI mode~\citep{Krolik2006} and vary from $2 \lambda_0$ to $16 \lambda_0$ ($\lambda_0$  is related to the plasma skin depth $d=c/\omega_p$ by $\lambda_0 = 2\pi (v_A/c)(\omega_p/\alpha) d$).
Time is normalized to the orbital period $P_0 = 2\pi/\alpha$. 
Other numerical parameters are summarized in Table \ref{tab:parameters}. The spacial resolution is chosen such as to simultaneously resolve both the skin depth $d$ and the Larmor radius $\rho$ (with $\rho=\sqrt{\beta}d$ and $\beta>1$) during the evolution of the simulation. 

\begin{table}[t]
	\centering
	\begin{tabular*}{\linewidth}{c|c|c|c|c}
		 & run A & run B & run C & run D \\
		\hline
		\hline
		$\beta$ & 100 & 100 & 100 & 100 \\
		$v_A/c$ ($\times$10$^{-2}$) & 1.43 & 1.43 & 1.43 & 1.43 \\
		$\Omega_0/\alpha$ & 11 & 11 & 11 & 11 \\
		$L_x$ = $L_z$ & 2 & 4 & 8 & 16   \\
		$N_x$ = $N_z$ & 552 & 1105 & 2210 & 4420 \\
		$\Delta_x$ = $\Delta_z$ [$c/\omega_p$] & 0.25  & 0.25 & 0.25  & 0.25 \\
		\# ppc & 25 & 25 & 25 & 25  
	\end{tabular*}
	\caption{Simulation parameters. The last three rows indicate, respectively, the total number of cells used in each direction, the numerical resolution, and the number of particles per cell used in the simulations.}
	\label{tab:parameters}
\end{table}

\section{Effect of simulation domain size} \label{sec:box_size}
\begin{figure}[th!]
	\centering
	\includegraphics[width=\columnwidth]{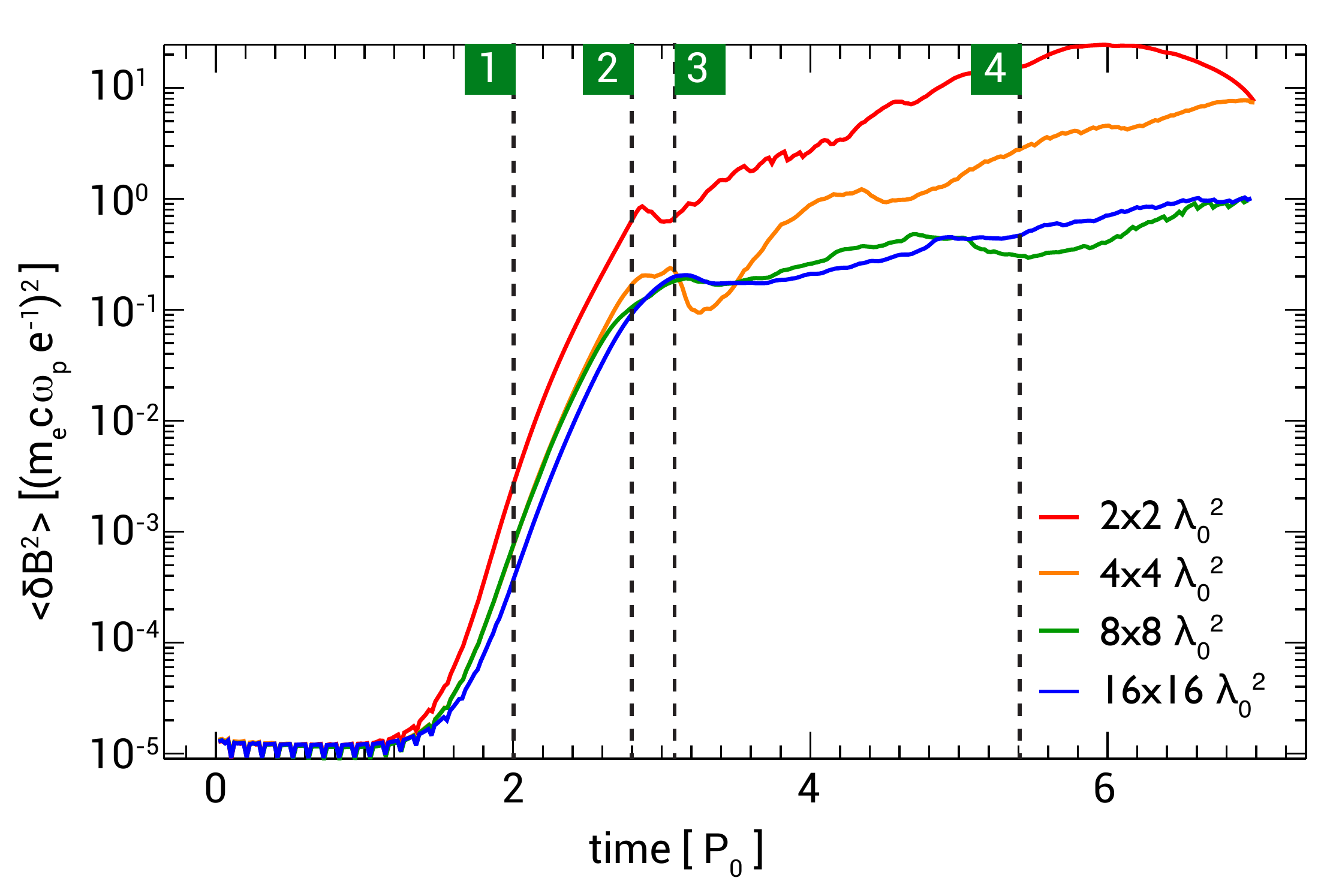}
	\caption{Time evolution of the domain-averaged magnetic energy fluctuation $<\delta B^2>$ for different simulation domains. The green dashed lines represent, respectively, the times $T_1=2 P_0$, $T_2=2.8 P_0$, $T_3=3.1 P_0$ and $T_4=5.4 P_0$.}
	\label{fig:ene_evolution}
\end{figure}

In this section, we analyze the time evolution of the collisionless MRI, with particular focus on two key ingredients of its nonlinear evolution: magnetic reconnection and pressure anisotropy generation. 

Fig.\ref{fig:ene_evolution} shows the time evolution of the domain-averaged magnetic energy fluctuation $<\delta B^2>$ for different simulation domains. 
We highlight four different phases that represent, respectively, the transition between the linear and the nonlinear regime (labeled `1'), and three meaningful stages of the nonlinear evolution that will be further analyzed below.

For simulation domains larger than $4\lambda_0$, we observe that our numerical simulations have converged. The noticeable differences in the nonlinear evolution between the small and the large simulation domains indicate that this is a key parameter in determining the dynamics, as will be further documented below.
The amplitude of the magnetic energy in the saturation regime of the instability decreases by almost one order of magnitude from small to large simulation domains until reaching convergence when the size of the simulation domain $L\geq8\lambda_0$.
\begin{figure}[h!]
	\centering
	\includegraphics[width=\columnwidth]{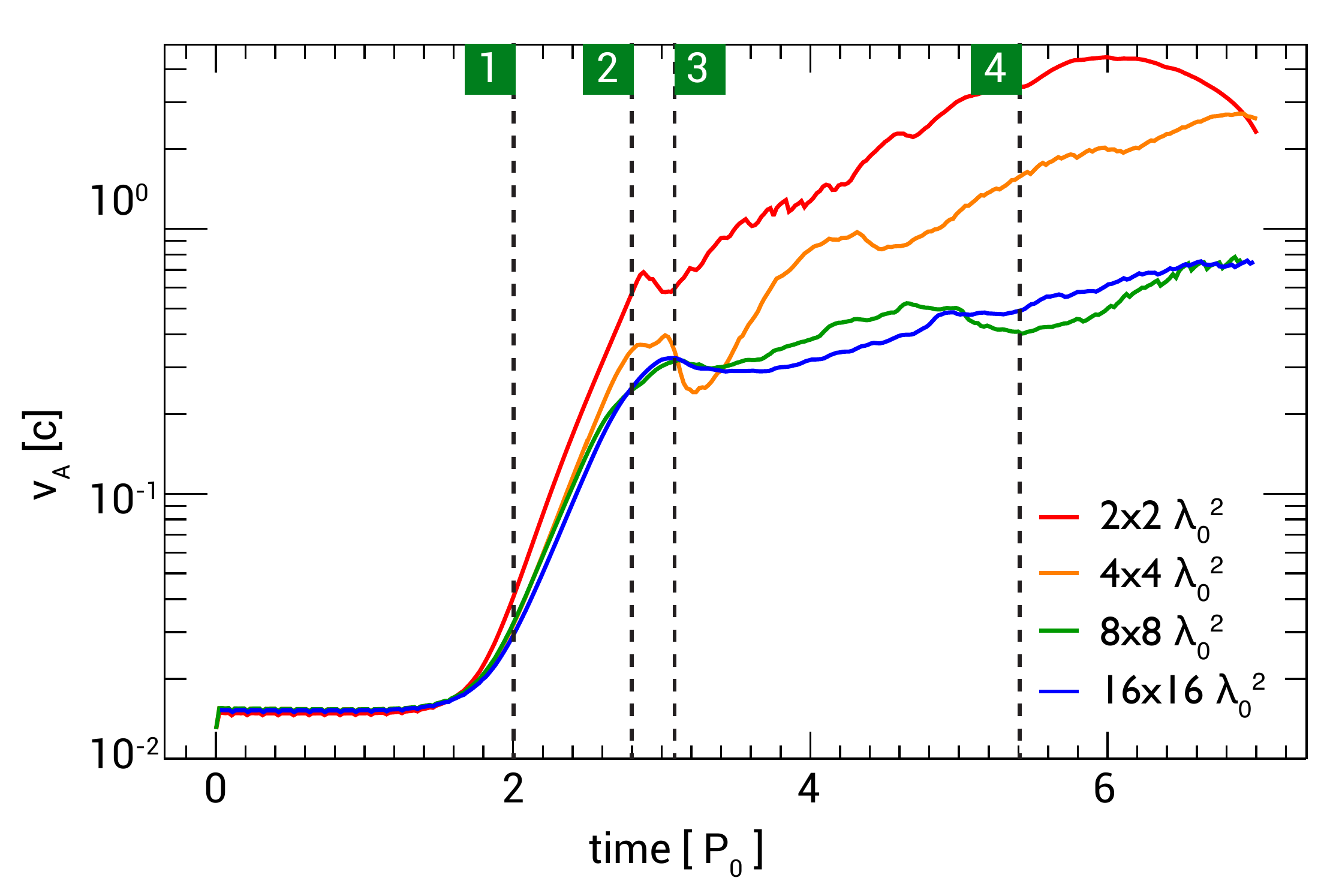}
	\caption{Time evolution of the domain-averaged Alfv\'en velocity $v_A$ for different simulation domains.}
	\label{fig:va_evolution}
\end{figure}

Figure \ref{fig:va_evolution} shows the time evolution of the domain-averaged Alfv\'en velocity $v_A$ for different simulation domains.
We observe that the evolution of $v_A$ depends on the simulation domain size: for small domains (run A), the Alfvén speed reaches values of $v_A\sim3c$, in agreement with \citep{Riquelme2012}. The conclusions for the small box case are then the same as those of \textit{Riquelme et al.}: for small domains, we do not observe a saturation of the MRI within the limit of validity of the shearing co-rotating framework. 
However, when the simulation domain size is increased, we observe that the exponential growth of the MRI saturates at values of the Alfv\'en velocity below the speed of light, within the validity range of our model; this is in agreement with our general observation that the size of simulation domain is critical to correctly capture the nonlinear dynamics of the collisionless MRI.
In addition, we checked \textit{a posteriori} the magnitude of the neglected terms in the Maxwell’s equations \ref{eq:gauss}-\ref{eq:ampere} compared to the terms that we keep, and found that, on average, those ratios are less than 1\% for the run D (16x16 box).

We will focus the analysis on our largest simulation, corresponding to a simulation domain of $16\lambda_0\times16 \lambda_0$,  run D in Table \ref{tab:parameters}.
\begin{figure}[t!]
	\centering
	\includegraphics[width=\columnwidth]{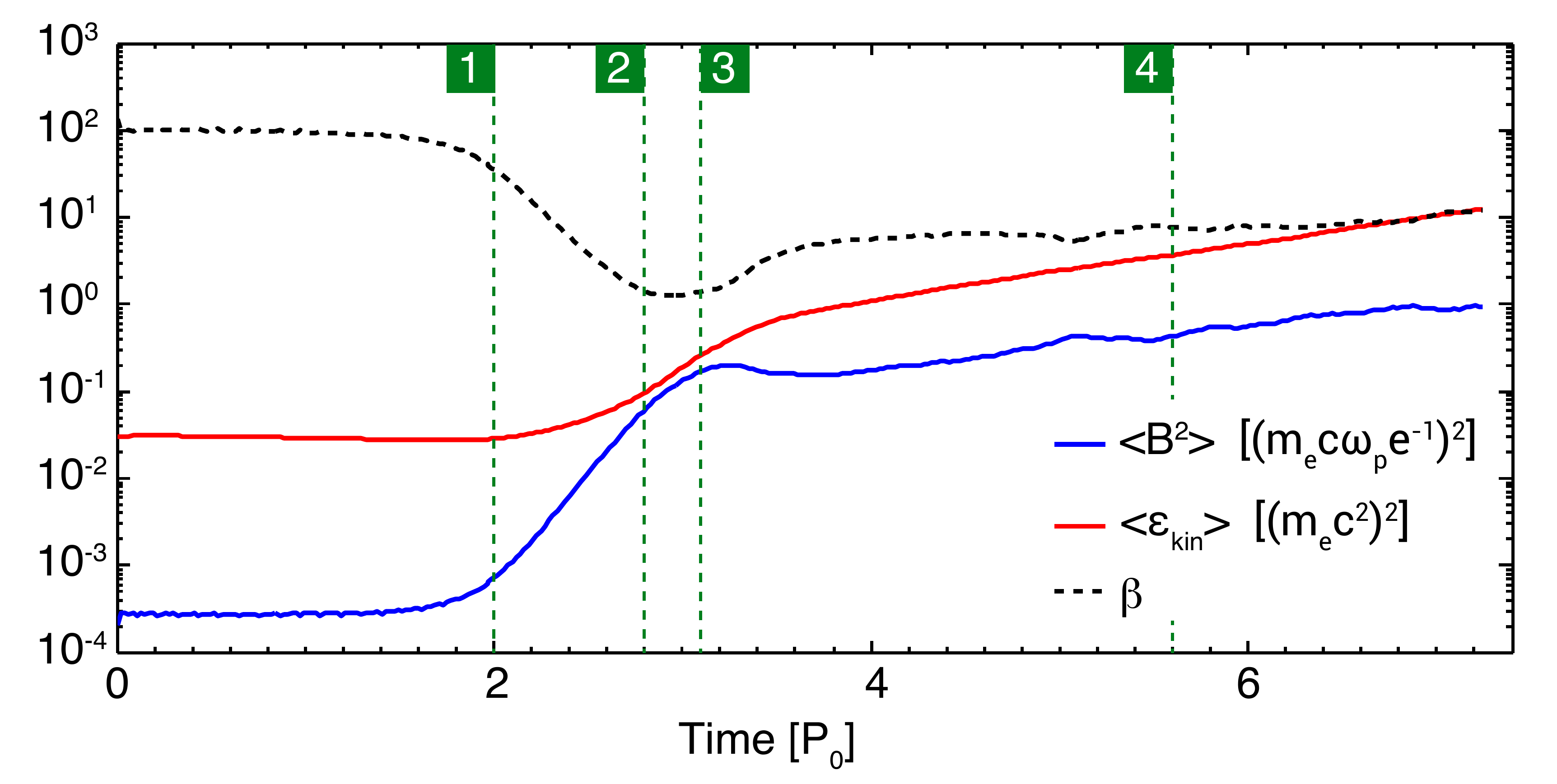}
	\caption{Time evolution of the domain-averaged magnetic energy (blue), kinetic energy (red) and $\beta$ parameter (dashed) for the $L=16 \lambda_0$ simulation (run D). The green dashed lines are at $T_1=2 P_0$, $T_2=2.8 P_0$, $T_3=3.1 P_0$ and $T_4=5.4 P_0$.}
	\label{fig:16_evolution}
\end{figure}

Figure \ref{fig:16_evolution} shows the time evolution of the domain-averaged magnetic and kinetic energy and the $\beta$ parameter for this run.
During the amplification of the magnetic field produced by the MRI, the $\beta$ parameter decreases from its starting value of 100, reaching the equipartition value at time $T_{orbit} = 3.1 P_0$. Interestingly, however, the value of $\beta$ does not remain around unity and increases up to 10 during the saturation regime of the instability.
\footnote{In figure \ref{fig:16_evolution} we observe that the beta parameter is practically constant throughout the late stages of evolution of the instability. However, both the magnetic and kinetic energy slowly rise during this period. In our setup, there is a continuous source of kinetic energy injection (the shearing term in our equations). The constancy of beta indicates that the system has reached an equilibrium between the kinetic and magnetic energies. However, in a true steady-state, the energy injected should match the energy dissipated, and thus both the kinetic and the magnetic energy should be constant (on average); this is not what we observe. It is possible that to attain a real steady state we would have to run the simulations for much longer, this would require significant computing resources that we currently do no have. Another possibility is that this secular growth is caused by insufficient energy dissipation in our code. Another possibility is that this is a manifestations of residual numerical effects such as numerical heating intrinsic to the PIC algorithm.}

As we will describe later, the magnetic reconnection that is activated along the phase 3 is the mechanism responsible for the growth of $\beta$, progressively converting the magnetic field energy into kinetic energy. 
This behavior is observed also in our smaller simulation domains (not shown here), and was also manifest in the previous work of \cite{Hoshino2013}. 

 \begin{figure*}[th!]
	\centering
	\includegraphics[width=\textwidth]{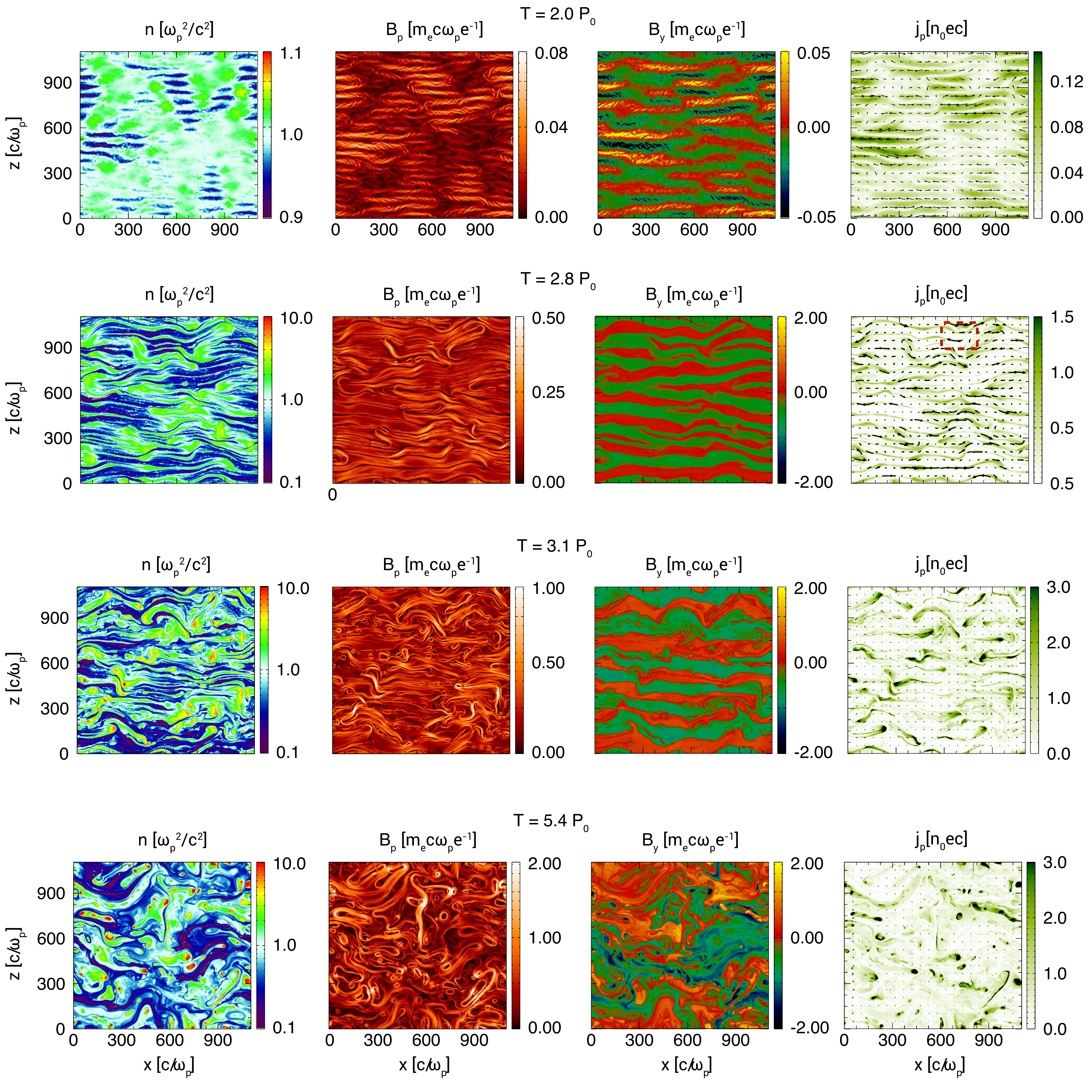}
	\caption{From left to right: average density $n=(n_++n_-)/2$, in-plane magnetic field $B_p = \sqrt{B_x^2+B_z^2}$, transverse magnetic field $B_y$ and in-plane  current density $j_p=\sqrt{j_x^2+j_z^2}$ at time $T_{orbit} = 2 P_0$ (top), $T_{orbit} = 2.8 P_0$ (middle up row), $T_{orbit} = 3.1 P_0$ (middle down row) and $T_{orbit} = 5.4 P_0$ (bottom row) for the $L=16 \lambda_0$ simulation (run D). The black arrow in the current plots indicate the current direction. The dashed square in the current plot represents the zoom in Fig. \ref{fig:current_zoom}}
	\label{fig:evolution}
\end{figure*}
In Fig. \ref{fig:evolution} we display snapshots of the average plasma density $n=(n_++n_-)/2$, the module of the in-plane magnetic field $B_p = \sqrt{B_x^2+B_z^2}$, the transverse magnetic field $B_y$ and the module of the in-plane current density $j_p=\sqrt{j_x^2+j_z^2}$ for run D at different times.
$T_{orbit} = 2 P_0$ represents the transition between the linear regime and the nonlinear regime of the MRI in the sense that the growing perturbation of the magnetic field $\delta B$ starts to be on the order of the initial external magnetic field $B_0$. 
This regime is characterized by the formation of coherent structures --- channel flows  \citep{Goodman1994} --- which, like the linear mode that they evolve from (see Fig.\ref{fig:disp_rel_appendix} in the Appendix), are on MHD scales. 
It is worth commenting in passing that this shows that channel flows are a robust MHD solution, still observed in the fully kinetic, collisionless regime that we explore here.

\begin{figure}[ht!]
	\centering
	\includegraphics[width=\columnwidth]{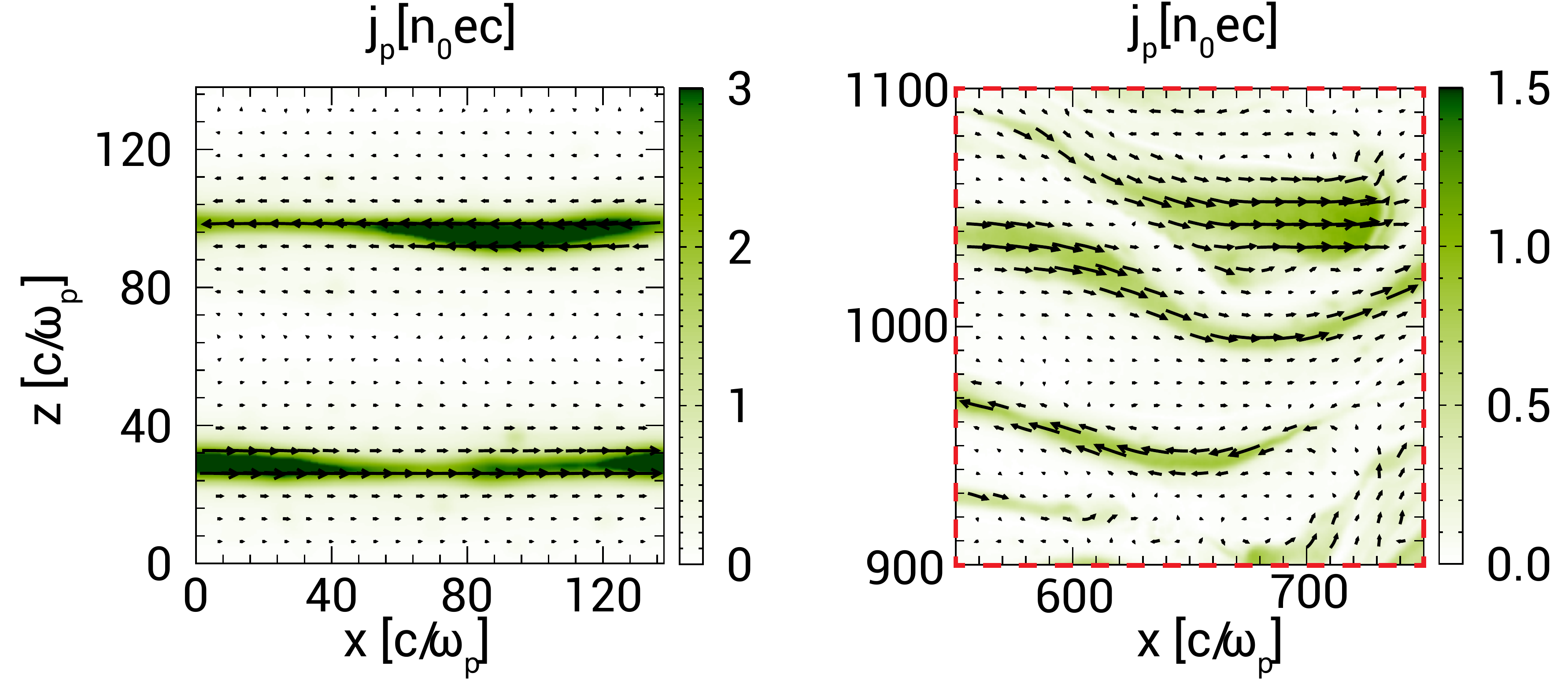}
	\caption{In-plane current density $j_p$ for the $L=2\times2\lambda_0^2$ (left) and a zoom of the $L=16\times16\lambda_0^2$ cases (right) at time $T_{orbit} = 2.8 P_0$. The black arrows indicate the current direction.}
	\label{fig:current_zoom}
\end{figure}

Channel flows are an exact solution of the nonlinear MHD equations \citep{Goodman1994}; as such, their amplitude would grow unbounded unless they are disrupted by parasitic instabilities.
We will now focus on how this occurs in run D.
The channel flows confine the plasma density between regions of positive and negative transverse magnetic field $B_y$, with the formation of current sheets in the plane of the simulation, as shown by the in-plane current density $j_p$ in Fig. \ref{fig:evolution}. 
From time $T_{orbit} = 2 P_0$ to $T_{orbit} = 2.8 P_0$, we observe that the thickness of the current sheets decreases with the growth of the transverse component $B_y$ of the magnetic field. 
These current sheets subsequently develop a radial modulation, as illustrated in Fig. \ref{fig:current_zoom}, right panel, which is a magnification of the red box identified in the plot of $j_p$ at time $T_{orbit} = 2.8 P_0$ in Fig. \ref{fig:evolution}.
For contrast, the left panel of Fig. \ref{fig:current_zoom} shows a snapshot of $j_p$ from our smallest simulation (run A) at the same time.
This comparison shows that the amplitude of this modulation is significantly more pronounced in the large simulation domain. 

For times $T_{orbit} > 2 P_0$, the plasma possesses well-defined current sheets along the radial direction that carry both $J_x$ and $J_y$ currents, with comparable magnitudes. 
These current sheets are surrounded by shearing magnetic fields $B_x$ and $B_y$,  with $B_x \sim B_y$. 
In principle, this configuration allows for the development of both tearing and drift-kink modes \citep{Pritchett1996, Daughton1998, Daughton1999, Daughton1999a} in the $x-z$ plane of the simulation, along the $x$ direction.\footnote{The drift kink investigations of \cite{Pritchett1996, Daughton1998, Daughton1999, Daughton1999a} consider a configuration where the only component of the magnetic field would correspond to our $B_x$. 
The channel flows whose stability we are discussing are threaded by both $B_x$ and $B_y$, with $B_x/B_y\sim 1$, and thus it is not immediately obvious that those results on the DK instability are still valid here. 
However, note that both $B_x$ and $B_y$ are modulated in the $z$-direction, and we will find that the DK instability is located at values of $z$ where $B_y\approx 0$, legitimizing our comparison with the aforementioned theories. } 

The characteristic thickness of the current sheets before the onset of the modulation can be measured from the simulation to be $\delta \sim (0.2-0.3)\lambda_0 \sim 20~c/\omega_p$ ---
see Fig. \ref{fig:current_zoom}. 
From the simulation, we can also measure the average wavenumber of the modulation to be approximately $k=2\pi/\lambda\sim 0.045~\omega_p/c$. 
These measurements are consistent with previous numerical studies of the drift-kink instability, where $k_{DK}\delta\sim 1$ is expected \citep{Zenitani2005, Daughton1999}. 

Importantly, observe that $\lambda_{DK}>\lambda_0$, which partially accounts for the need to have large simulation domains.
In particular, we notice that the amplification of the DK modulation is larger for larger domains.
For small domains, the growth of the magnetic field proceeds unhindered until the current sheets become unstable to reconnection; in such cases, modulation of the current sheet due to the DKI is small or non-existent. 
This has important consequences for the subsequent nonlinear dynamics. 
In small domains, where DKI is mostly absent, the motion of the magnetic islands formed once the channel flows break is mostly confined in the radial ($x$) direction of the simulation. 
When the size of the simulation domain is increased, the different current sheet structures and the larger amplitude of the DK modes activate a non-uniform motion of the current sheets along the vertical ($z$), as well as radial, directions. Additionally, there is a much larger variety of island sizes produced. The combination of these different effects results in a transition to fully turbulent dynamics that is absent in smaller simulation domains.

\section{Pressure anisotropy driven instability}\label{sec:mirror}
\begin{figure}[ht!]
	\centering
	\includegraphics[width=\columnwidth]{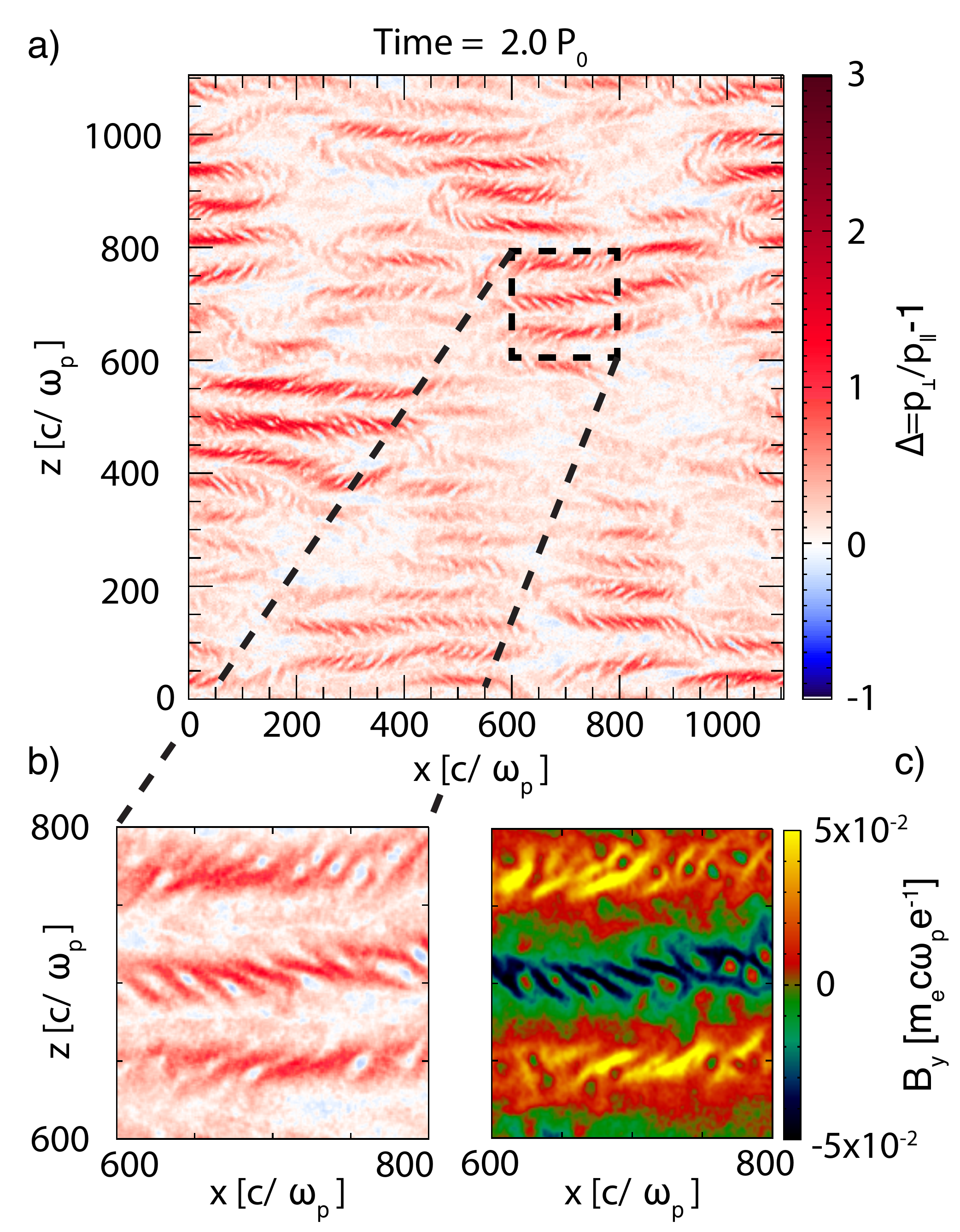}
	\caption{a) Pressure anisotropy $\Delta = p_\perp/p_\parallel-1$ at $T_{orbit} = 2 P_0$. A magnified section of the simulation domain is shown in b), with the corresponding magnetic field plotted in c). Oblique magnetic field structure filaments form in the regions of maximum anisotropy. }
	\label{fig:anisotropy}
\end{figure}
Fig. \ref{fig:anisotropy} shows the spatial distribution of the pressure anisotropy at time $T_{orbit} = 2 P_0$, represented by the parameter $\Delta = p_\perp/p_\parallel-1$. 
The pressure anisotropy grows with the growth of the magnetic field induced by the MRI. 
In particular, the regions of maximum anisotropy are where the magnetic field forms filaments oblique to the direction of the external magnetic field, as shown in Fig. \ref{fig:anisotropy} b) and c). 
These results are consistent with previous numerical studies of the mirror instability \citep{Riquelme2012, Hoshino2013, Kunz2014}.
\begin{figure}[ht!]
	\centering
	\includegraphics[width=\columnwidth]{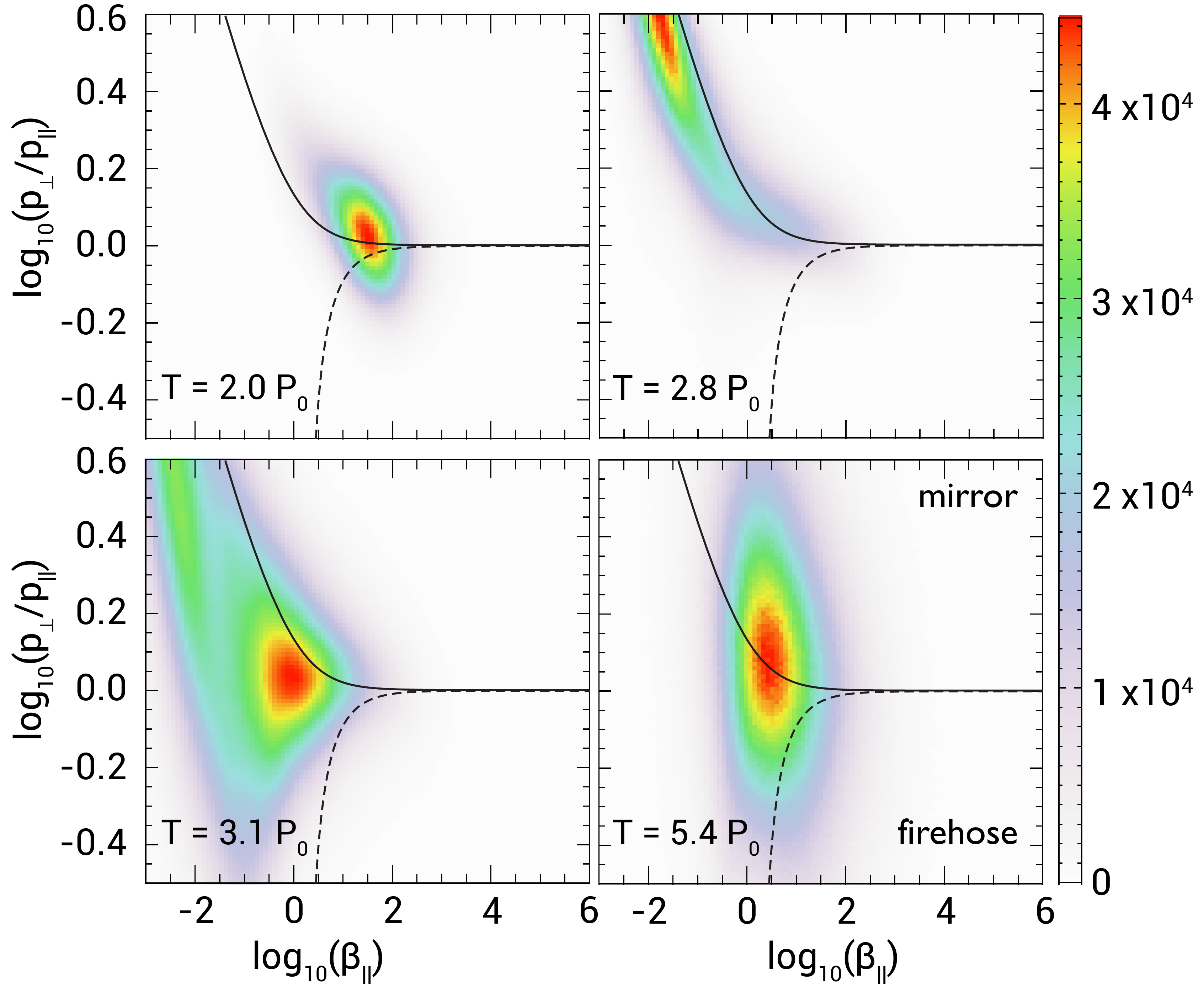}
	\caption{Distribution of the pressure anisotropy $p_\perp/p_\parallel$ as a function of the parallel $\beta_\parallel$ for different times. The solid line denotes the region where the plasma is mirror unstable (top) and the dashed line the region where it is firehose unstable (bottom).}
	\label{fig:anisotropy_distribution}
\end{figure}
Figure \ref{fig:anisotropy_distribution} shows the distribution of the pressure anisotropy $p_\perp/p_\parallel$ as a function of $\beta_\parallel = 8\pi p_\parallel/B^2$, where $B$ is the total magnetic field, at different times. 
The solid line marks the stability threshold of the mirror instability, $p_\perp/p_\parallel -1/2 > \sqrt{1/4+1/2\beta_\parallel}$~\citep{Pokhotelov2000}.
The dashed line denotes the stability threshold of the firehose instability,  $p_\perp/p_\parallel -1 < 2/\beta_\parallel$ \authorcomment2{there was a mistake in the threshold, missing a factor 2}\citep{Yoon1993, Hellinger2000}.
At time $T_{orbit} = 2 P_0$, the distribution of the pressure anisotropy is mostly above the threshold of the mirror instability. 
The combined action of further magnetic field amplification due to the MRI and of the mirror instability moves the pressure anisotropy to regions of lower $\beta_\parallel$ and within the stability margins, as obtained at $T_{orbit}=2.8P_0$. 
At this time, there is thus a balance between pressure anisotropy generation by the MRI, and its destruction by the mirror instability \citep{Kunz2014, Riquelme2015, Melville2016}.

Note that in a realistic disc, one expects $\alpha/\Omega_0 \ll 1/\beta$ ; the growth rate of the mirror instability would therefore always be much larger than that of the MRI, implying that the signature of the mirror instability should appear earlier in time than what we obtain in our simulations.
Due to numerical constraints, this condition is not verified with the initial parameters of our simulations. During the $\mu$-conserving phase of the MRI, however, the growth of the magnetic field simultaneously reduces the $\beta$ parameter and increases $\Omega_0$, such that the above condition becomes verified. This delay of the effects of the mirror instability explains why its saturation only occurs at $\sim2.8$ orbits.

When the MRI starts to saturate at $T_{orbit} = 3.1 P_0$, the distribution of the pressure anisotropy is completely within the stability bounds. 
However, an interesting feature emerges at this stage:  a secondary peak of the distribution arises at $p_\perp/p_\parallel\sim 1$.
At time $T_{orbit} = 5.4 P_0$, this peak has grown, and we observe strong violation of the mirror stability boundary at $\beta_\parallel\sim3$. 
As discussed in Sec. \ref{sec:box_size}, at this time the motion of the plasma is turbulent and magnetic reconnection plays a crucial role in the dynamics and coalescence of plasma islands. 
We think reconnection is the origin of this violation of the mirror stability threshold, as we now explain.

In two-dimensional geometry, previous numerical studies of reconnection \citep{Zenitani2001, Sironi2014, Dahlin2016} show that the particles are accelerated mostly along the direction perpendicular to the reconnection plane ($y$ in our configuration).
The typical duration of a reconnection event in our simulations is the Alfv\'en time $\tau_A=l/v_A$, where $l$ is the length of a given current sheet and $v_A$ the local Alfv\'en velocity.
Along the direction perpendicular to the plane of reconnection, the dominant contribution to the acceleration of a particle is due to the reconnection  electric field. 
The amplitude of this electric field is estimated to be (in normalized units) $E_{y}\sim0.1v_A/cB$ \citep{Lyubarsky2005}, where the factor $0.1$ represents the characteristic relativistic reconnection rate \citep{Liu2015,Cassak2017}. 
With this assumption, we can estimate the velocity gain along  $y$ to be $\Delta(\gamma v_y)/\tau_A\sim e/m E_y$, where $\gamma$ is the Lorentz factor. 
This yields a final proper velocity of  $u_y=\gamma v_y\sim 0.1v_A l/d$, where $d=c/\omega_p$ is the skin depth.
In our simulations, the typical length of the current sheet is $l\gtrsim100~c/\omega_p$ and thus, on average, we expect $v_y \approx 10 v_A$.
In the reconnection plane (i.e., the plane of the simulation), instead, reconnection-accelerated particles typically move at the proper Alfv\'en velocity $v_x\sim v_A$ \citep{Lyubarsky2005}. 
 This suggests a mechanism for the generation of velocity anisotropy, with $v_y>v_x$. 

With the above estimates, one can predict the growth time for the mirror instability resulting from this reconnection-generated pressure anisotropy to be \cite{Pokhotelov2000} $\tau_M\sim(1+(l/10d)^2)\tau_A\sim 100\tau_A$. 
This is much longer than the typical reconnection event, explaining, we believe, why the mirror stability boundaries are violated at this stage of our simulations.
 
In this regard, our results are different from \cite{Kunz2016}, where, instead, the mirror instability threshold remains a solid boundary constraining the nonlinear dynamics. 
It is conceivable that the differences between our results and theirs stem from additional constraints imposed by the two-dimensional geometry that we use; but, alternatively, it is also possible that  we have uncovered an effect that critically depends on a kinetic treatment of electrons, which Kunz et al. (2016) do not do.
The extension of our work to fully three-dimensional geometries requires extraordinary computational resources and must thus be left for future work.

\section{MRI turbulence}\label{sec:turbo}
\begin{figure}[ht!]
	\centering
	\includegraphics[width=0.86\columnwidth]{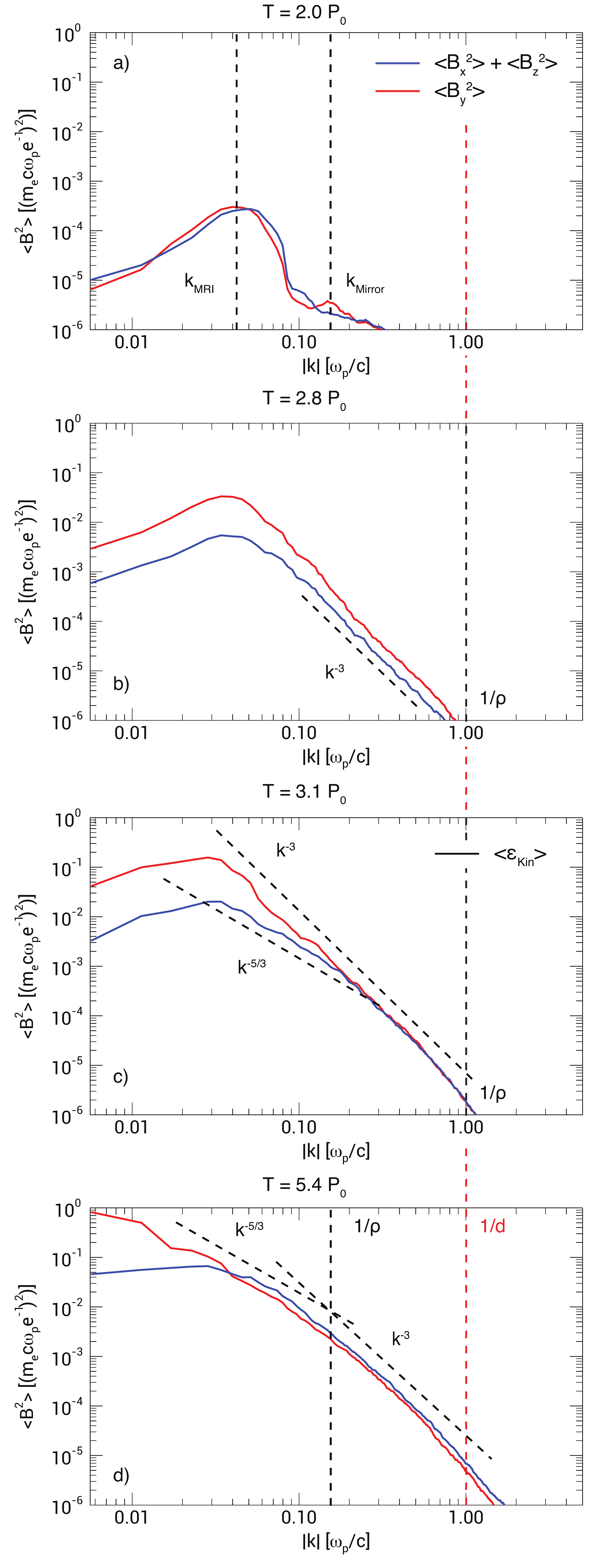}
	\caption{Domain-averaged magnetic energy spectrum for the in-plane ($<B_p>$, red) and the transverse ($<B_y>$, blue) components of the magnetic field at different times.
	The dashed red line represents the plasma skin depth $1/d = \omega_p/c$. The black dashed lines represent, respectively, the maximum wavelength of MRI and the maximum wavelength of the mirror instability  in the subplot a) and the Larmor radius $1/\rho$ in subplots b), c), d). }
	\label{fig:turbulence_all}
\end{figure}
Figure \ref{fig:turbulence_all} shows the magnetic energy spectrum for both the in-plane $B_p$ and transverse $B_y$ components at different times.
The energy spectrum is defined as $<B_j^2> = \int d\Omega_k(k/2\pi)^2|B_j|^2$, where $k=\sqrt{k_x^2+k_z^2}$ and $\Omega_k = \tan^{-1}(k_x/k_z)$.
During the linear regime (subplot a) at time $T_{orbit} = 2.0 P_0$, the energy spectrum shows a peak at $k_{MRI}\sim0.04~\omega_p/c$, that corresponds to the maximum wavelength of the collisionless MRI in our system (see appendix \ref{sec:appendix}). 
In the in-plane $B_p$ energy spectrum, we can also observe a secondary peak at $k_{Mirror}\sim0.15~\omega_p/c$ which is consistent with the maximum wavelength of the mirror instability \citep{Pokhotelov2000}. This coexistence of the MRI and the mirror instability is clearly visible in Figure \ref{fig:anisotropy}.

At $T_{orbit} = 2.8 P_0$, the energy spectrum shows a well-defined power law distribution for high $k$, with a $-3$ slope for both the in-plane and transverse component of the magnetic field. 
This slope occurs at $k\rho \ll1$ (recall from Figure \ref{fig:16_evolution} that $\beta\approx 1$ at this time, so $d\approx \rho$ and thus also $k d\ll 1$).
It is not obvious why we observe this power-law behavior at this stage, since it occurs before the transition to fully-developed MRI turbulence. 
It could conceivably be the result of mirror instability-driven turbulence, except that that is predicted to yield a $-5/3$ slope at the fluid scales \citep{Kunz2014}, different from what we observe.
A tentative explanation is that this power-law behavior is due to the effect of magnetic reconnection at those scales, as suggested by recent analytical predictions \citep{Loureiro2017b, Mallet2017} (with the caveat that these predictions were made for ion-electron plasmas, not pair plasmas). Visual evidence for magnetic reconnection occurring already at this time is discernible in the contour plots of Figure \ref{fig:evolution}, second row. 

At time $T_{orbit} = 3.1 P_0$, the role of reconnection in the nonlinear dynamic of the system is more pronounced, with the disruption of the MRI channel flows and activation of large scale turbulence.
The transverse component of the magnetic energy spectrum maintains the slope of $-3$ at fluid scales.
The in-plane component of the magnetic field, instead, shows a transition from a slope consistent with the familiar $-5/3$ at large scales, to $-3$.

At time $T_{orbit} = 5.4 P_0$, the plasma is in a fully turbulent state, as shown in the last row of Fig. \ref{fig:evolution}. The last subplot in Fig \ref{fig:turbulence_all} shows that the spectrum has a characteristic slope of $k^{-5/3}$ for scales bigger than the Larmor radius ($k\rho <1$), and retains the $-3$ slope at sub-Larmor scales. Such a slope in the kinetic range is consistent with expectations from kinetic Alfv\'en (or perhaps whistler) wave turbulence \citep{Howes2008, Schekochihin2009, Chen2010, Boldyrev2012, Passot2015}, but also with reconnection-mediated turbulence ~\citep{Loureiro2017b, Cerri2017, Franci2017}.

\section{Particle acceleration}\label{sec:acceleration}
\begin{figure}[ht!]
	\centering
	\includegraphics[width=\columnwidth]{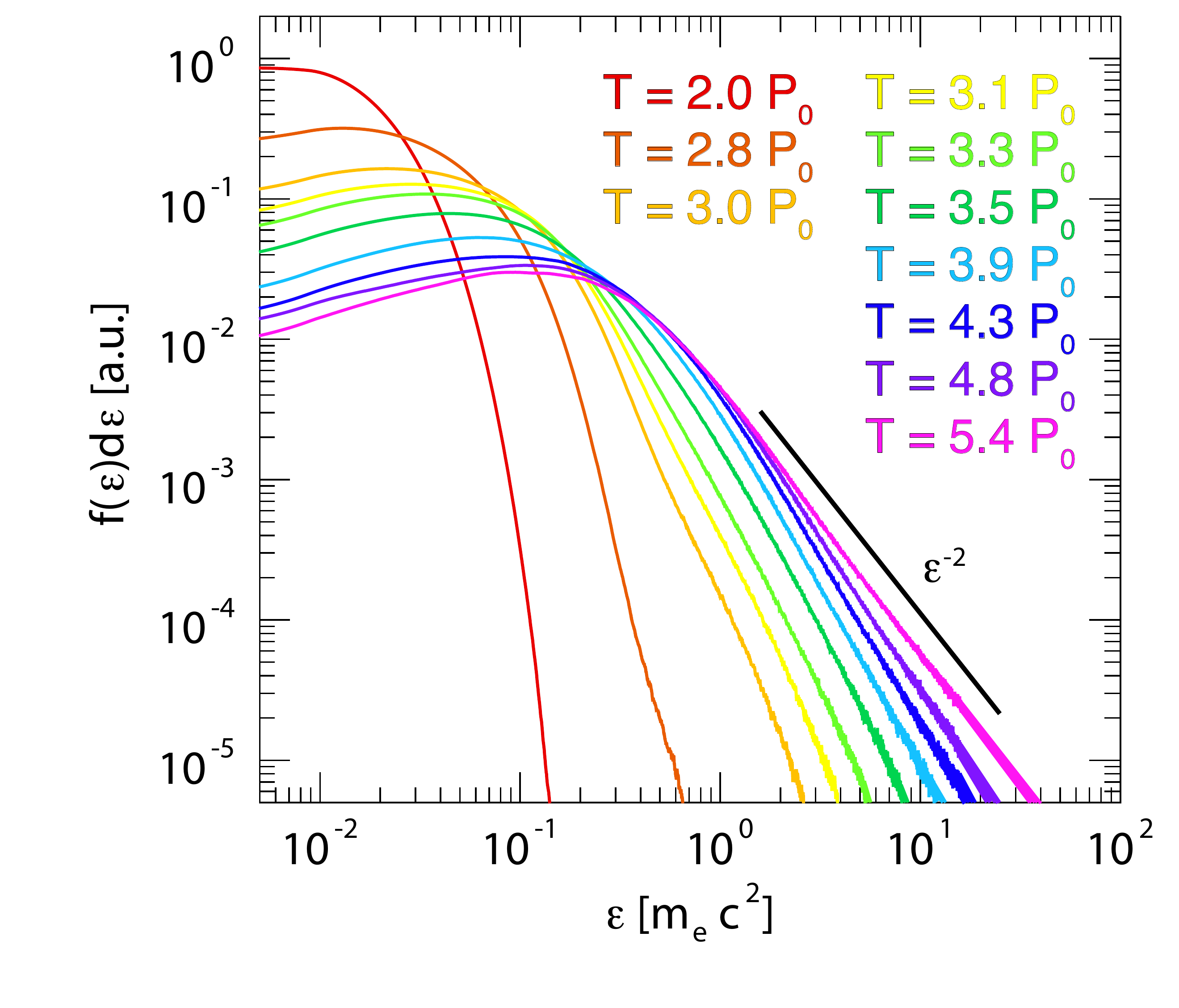}
	\caption{Evolution of kinetic energy distribution for the simulation $L=16\lambda_0$. Before reconnection, the distribution remains hot Maxwellian-like ($T_{orbit} = 2 P_0$). After reconnection, high-energy, non thermal particles are generated. At the steady state regime ($T_{orbit} = 5.4 P_0$) the high energy component of the distribution can be approximated by a power-law function  with $f(\epsilon)d\epsilon\propto\epsilon^{-2}$.}
	\label{fig:energy_distribution}
\end{figure}
In Fig. \ref{fig:energy_distribution} we plot the evolution of the kinetic energy distribution at different times. 
At $T_{orbit} = 2 P_0$, corresponding to the late linear evolution of the MRI, the energy distribution is still thermal. 
At times $T_{orbit} = 2.8 P_0$ and $T_{orbit} = 3 P_0$, corresponding to the activation of reconnection, we observe the development of a hot tail in the energy distribution, consistent with the claim made in Section \ref{sec:mirror} that reconnection is efficient at the generation of high-energy particles. 
At late times, during the steady state turbulence regime, we see that the kinetic energy distribution has evolved to exhibit a clear power law slope of $-2$ at high energies.
Recent numerical studies of particle acceleration via relativistic magnetic reconnection \citep{Werner2015} show that the slope of the particle acceleration depends of the ratio between the the typical length of the current sheets $l$ and the magnetization parameter $\sigma$. For a hot plasma, as is the case here, the magnetization parameter can be defined as $\sigma = B^2/(4\pi n(\gamma m_ec^2+2.5T))$, where $T$ is the plasma temperature.\citep{Melzani2013}
Using for these parameters the values obtained in our simulations close to current sheets ($l\sim100-200 c/\omega_p$ and $\sigma\sim3-5$), the predicted slope can vary between $-2$ and $-2.5$ \citep{Werner2015}, in good agreement with our simulations.

The role of reconnection in generating non-thermal tails in the nonlinear development of the MRI has been previously pointed out by \cite{Hoshino2013, Hoshino2015} who, however, obtained a shallower slope of $-1$, possibly due to the smaller simulation domain employed there.

\section{Summary}\label{sec:conclusion}
In this work, we numerically analyzed the fully kinetic nonlinear evolution of two-dimensional MRI in a collisionless, high $\beta$ pair plasma. 
Our main results are the following. 
(i) The amplitude of the channel flows generated in the early nonlinear regime is limited by the onset of the drift-kink instability (DKI), and subsequent magnetic reconnection. 
Both play a critical role in the transition to a regime of fully-developed turbulence. 
Importantly, we observe that simulations with smaller domain sizes yield insignificant DKI; as a result, such simulations fail to develop substantive turbulent dynamics.
One of our noteworthy conclusions, therefore, is that sufficiently large simulation domains (compared to the wavelength of the most unstable MRI mode) are required to reach saturation.\footnote{This conclusion does not explicitly contradict any previous numerical study that we are familiar with. The two-dimensional kinetic MHD simulations reported in \cite{Riquelme2012} fail to saturate, but they are performed in simulation domains of the order of  $\sim\lambda_0$. For such a domain, our simulations also do not saturate.}
(ii) Reconnection leads to significant velocity anisotropy. It is more efficient at generating such anisotropy than the mirror instability is at destroying it. 
As a result, the nonlinear turbulent plasma state significantly violates the mirror stability boundary.  
(iii) During the initial phase of the nonlinear MRI, the magnetic energy spectrum presents a characteristic slope of $-3$ for scales larger that the Larmor radius.
We interpreted this result as the generation of a turbulent regime at these scales driven by magnetic reconnection.
Subsequently, the turbulence is activated also at larger scale and the magnetic energy spectrum presents a $-5/3$ slope for scales larger than the Larmor radius and a slope $-3$ for sub-Larmor scales.
(iv) A energetic particle spectrum is obtained, well described at high energies by a power-law with slope $-2$.
 
Two obvious limitations of our work are the reduced mass ratio employed (we consider a pair plasma) and the dimensionality (2D instead of 3D). Going forward, we aim to address both of these issues to assess the extent to which they affect the conclusions drawn here.
In particular, pressure-anisotropy generation by magnetic reconnection may be strongly impacted by the dimensionality of the setup, given that in 3D we expect the current sheets to be oriented obliquely to the $x-z$ plane that we simulate here. 
In addition, the introduction of a significant mass ratio between the two plasma species may lead to interesting multiscale effects, such as the generation of pressure anisotropy-driven instabilities both at electrons and ions scales, that are absent from our simulations and cannot be captured via hybrid simulations.

At a qualitative level, the MRI dynamics evidenced by our simulations resembles that observed in MHD studies: the linear growth period is followed by a stage of channel flow formation; these channel flows disrupt due to parasitic instabilities and lead to a fully turbulent saturated state. Quantitatively, however, there are differences: for example, in the way that kinetic instabilities (i.e. mirror, drift kink) regulate certain stages of the evolution; the fact that our dominant parasitic mode seems to be magnetic reconnection instead of the usual Kelvin-Helmholtz of MHD, and how this is intimately linked to efficient particle acceleration; and, of course the details of the energy spectrum itself. Whether these differences matter in terms of transport and other macroscopic properties of the system remains to be understood.

This work was supported in part by the Funda\c{c}\~ao para a Ci\^encia e a Tecnologia (FCT) under the grant PD/BD/105855/2014 and by the European Research Council under the Advanced Grant In-Pairs n.695088.
We would like to acknowledge the assistance of high performance computing resources (Tier-0) provided by PRACE on Marenostrum4 based in Spain.
Simulations were performed at the IST cluster (Lisbon, Portugal), and the Marenostrum4 supercomputer (Barcelona, Spain).
NFL was supported by NSF CAREER award no. 1654168.
GI acknowledges M. W. Kunz, M. Hoshino and C. Ruyer for valuable discussions.

\appendix

\section{Linear regime of MRI}\label{sec:appendix}
In this appendix, we derive a one-dimensional two-fluid model for the linear regime of the MRI in the shearing co-rotating framework for pure vertical initial magnetic field\footnote{This can be generalized to include a finite azimuthal field $B_y$ \citep{Quataert2002}; in that case, the dispersion relation becomes sensitive to kinetic effects and would provide a complementary test of our algorithm. Here we adopt this setup because it is the one used for our simulations.}, and use the analytical results as a benchmark of our numerical code. 

In the non-relativistic limit of the shearing co-rotating frame,  Faraday's and Amp\'ere's equations are (see Eq. \ref{eq:three-dimensionshear_faraday} - \ref{eq:three-dimensionshear_ampere}):
\begin{eqnarray}
	 \frac{\partial\bm{B}}{\partial t} & = & -c\nabla\times\bm{E} - \frac{3}{2} \alpha B_x\hat{y} -\frac{3}{2}\alpha ct\frac{\partial\bm{E}}{\partial y}\times\hat{x}, \label{eq:three-dimensionshear_faraday_appendix}\\
    \frac{\partial\bm{E}}{\partial t} & = & c\nabla\times\bm{B}-4\pi\bm{J} -\frac{3}{2}\alpha E_x\hat{y}  -\frac{3}{2}\alpha ct    \frac{\partial\bm{B}}{\partial y}\times\hat{x}. \label{eq:three-dimensionshear_ampere_appendix}
\end{eqnarray}
\begin{figure}[t!]
	\centering
	\includegraphics[width=0.46\textwidth]{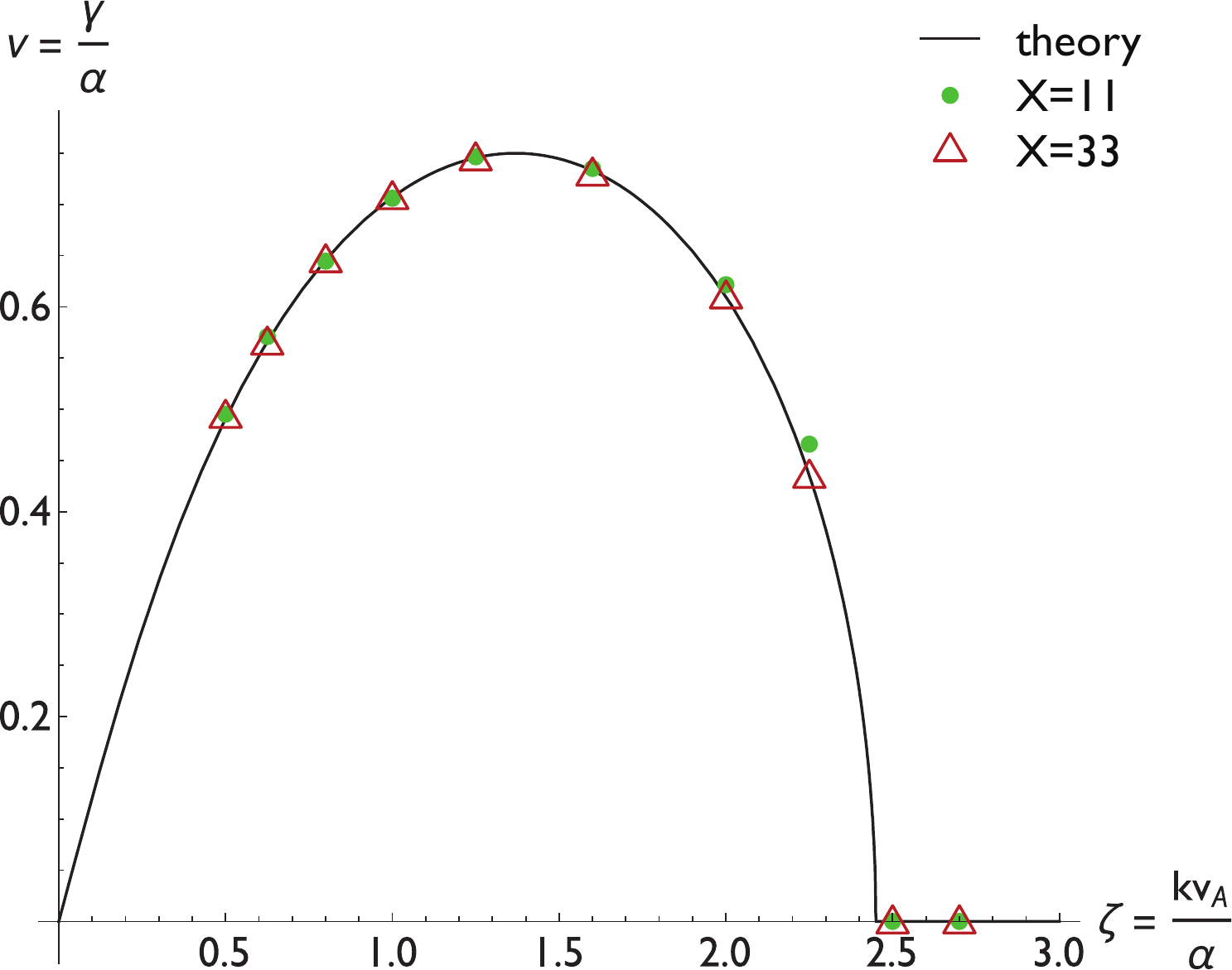}
	\caption{Analytical one-dimensional linear dispersion relation of MRI (black line) and numerical results for $X=11$ (green dots) and $X=33$ (red triangles).}
	\label{fig:disp_rel_appendix}
\end{figure}
In the limit that all fluctuations have wavelengths much larger than the ion Larmor radius and frequencies much smaller than the ion cyclotron frequency, the momentum and the continuity equations become:
\begin{eqnarray}
	m_jn_j\frac{\partial\bm{v}_j}{\partial t} = 2\alpha m_jn_jv&_{z,j}\hat{y} - \frac{1}{2}\alpha m_jn_jv_{y,j}\hat{z} + q_jn_j(\bm{E} + \frac{\bm{v_j}}{c}\times\bm{B}), \label{eq:moton_appendix}\\
	\frac{\partial n_j}{\partial t} + \nabla\cdot(n_j\bm{v_j}) =0,& \label{eq:continuity_appendix}
\end{eqnarray}
where the suffix $j$ indicates both the particles species (electrons and ions). 
The current is computed from Amp\'ere's law as:
\begin{equation}
	\bm{J} = \sum_j q_jn_j\bm{v}_j\label{eq:current_appendix}.
\end{equation}

We consider an external magnetic field $\bm{B} = B_0\hat{z}$ and a quasi-neutral equilibrium ($n_{0,e}\approx n_{0,i} \equiv n_0$). 
Linearizing Eq. (\ref{eq:three-dimensionshear_faraday_appendix} - \ref{eq:current_appendix}) and seeking solutions of the form $\exp(\gamma t + ikz)$, 
 we obtain the one-dimensional linear dispersion relation of MRI in the limit of weak magnetic field ($v_A\rightarrow0$), low rotational frequency ($X \gg 1$), cold ($\beta = 0$) and pair plasma ($R = 1$):
\begin{equation}
	4\nu^4+4(1+\zeta^2)\nu^2+\zeta^4-6\zeta^2=0,\label{eq:disp_rel_appendix}
\end{equation}
where we have adopted the following normalization:
\begin{eqnarray}
	\nu=\frac{\gamma}{\alpha},\hspace{2em} X=\frac{\omega_{ci}}{\alpha},\hspace{2em} R=\frac{m_i}{m_e}, \hspace{2em} v_A=\frac{\omega_{ci}}{\omega_{pi}},\hspace{2em}\zeta=\frac{kv_A}{\alpha}.
\end{eqnarray}
Eq. \ref{eq:disp_rel_appendix} is the same as obtained in \cite{Krolik2006} in the limit of pair plasma.

To verify the validity of our numerical shearing co-rotating framework, we consider a pair plasma with $\beta = 0.05$. The external magnetic field is defined by the Alfvén velocity $v_A = 0.05c$ and the angular frequency $\alpha$ by two different values of $X=11,\, 33$. 
The grid resolution of the numerical simulation is set to $\Delta x = 0.007 c/\omega_{pe}$ with 1000 particles per cell. 
To force the excitation of a specific MRI wavelength in our simulation, we seed the plasma with a velocity profile $\bm{v}_{seed}/c = 1/20\, v_A/c\,\sin(2\pi z/L)\hat{y}$, where $L$ is the size of the domain, scanned over the values $L=0.37, 0.4, 0.44, 0.5, 0.625, 0.8, 1.0, 1.25, 1.6, 2.0$. 

Figure \ref{fig:disp_rel_appendix} shows very good agreement between Eq. (\ref{eq:disp_rel_appendix}) and the numerical results obtained with the one-dimensional version of the shearing co-rotating version of the PIC code OSIRIS that we developed for this work.

\bibliographystyle{aasjournal}
\bibliography{./Bibliography.bib}

\end{document}